\documentclass{aastex62}
\usepackage{graphicx}
\usepackage{subfigure}
\usepackage{epstopdf}
\usepackage{ctable}
\usepackage{url}
\usepackage{times}
\usepackage{makecell}
\usepackage{tablefootnote}

\newcommand{\lsun}{~\mathrm{L_{\odot}}}

\newcommand{\msun}{~\mathrm{M_{\odot}}}
\newcommand{\msunperyr}{~\mathrm{M_{\odot} {\rm ~yr}^{-1}}}

\newcommand\fref[1]{\hyperref[#1]{Fig.~\ref*{#1}}}

\shorttitle{[CII] companions around $z>6$ galaxies}
\shortauthors{Miller et al.}

\received{XXX}
\revised{YYY}
\accepted{ZZZ}
\submitjournal{ApJ}

\begin{document}

\title{Investigating overdensities around $z>6$ galaxies through ALMA observations of [CII]}
\correspondingauthor{Tim B. Miller}
\email{tim.miller@yale.edu}

\author{Tim B. Miller}
\affiliation{Department of Physics and Atmospheric Science, Dalhousie University, 6310 Coburg Road, Halifax, NS B3H 4R2, Canada}
\affiliation{Department of Astronomy, Yale University, 52 Hillhouse Avenue, New Haven, CT, USA, 06511}

\author{Scott C. Chapman}
\affiliation{Department of Physics and Atmospheric Science, Dalhousie University, 6310 Coburg Road, Halifax, NS B3H 4R2, Canada}
\affiliation{NRC Herzeberg, 5071, West Saanich Rd, Victoria, BC V9E 2E7, Canada}

\author{Christopher C. Hayward}
\affiliation{Center for Computational Astrophysics, Flatiron Institute, 162 Fifth Avenue, New York, NY 10010, USA}

\author{Peter S. Behroozi}
\affiliation{Department of Astronomy and Steward Observatory, University of Arizona, Tucson, AZ 85721, USA}

\author{Matt Bradford}
\affiliation{California Institute of Technology, 1200 E. California Boulevard, Pasadena, CA 91125, USA}
\affiliation{Jet Propulsion Laboratory, 4800, Oak Grove Dr., Pasadena, CA 91109, USA}

\author{Chris J. Willott}
\affiliation{NRC Herzeberg, 5071, West Saanich Rd, Victoira, BC V9E 2E7, Canada}

\author{Jeff Wagg}
\affiliation{Square Kilometre Array Organization, Jodrell Bank Observatory, Lower Withington, Macclesfield, Cheshire SK11 9DL, UK Ran}

\begin{abstract}

We present a search for companion [CII] emitters to known luminous sources at $6< z <6.5$ in deep, archival ALMA observations. The observations are deep enough to detect sources with L$_{\rm [CII]} \sim 10^8\ L_\odot$ at $z \sim6$. We identify 3 new robust line detections from a blind search of five deep fields centered on ultra-luminous infrared galaxies and QSOs. We calculate the volume density of companions and find a relative over density of $6^{+4}_{-3}$ and $86^{+60}_{-37}$ when comparing to current observational constraints and theoretical predictions, respectively. These results suggest that the central sources may be highly biased tracers of mass in the early Universe. We find these companion lines to have comparable properties to other known galaxies at the same epoch. All companions lie less than  650\,km s$^{-1}$ and between 25 -- 60\,kpc (projected) from their central source. To place these discoveries in context, we employ a mock galaxy catalog to estimate the luminosity function for [CII] during reionization and compare to our observations. The simulations support this result by showing a similar level of elevated counts found around such luminous [CII] sources.
\end{abstract}


\section{Introduction}

In order to further our understanding of galaxy formation we must investigate how the first galaxies formed during the epoch of reionization (EoR). Advances in (sub)-millimeter interferometers have made it possible to detect galaxies out to a redshift of 6 and beyond both in continuum and spectroscopically ~\citep{Riechers2013, Maiolino2015, Strandet2017,Marrone2018}.This enables constraints on their physical properties such as star formation rate (SFR), dynamical mass and conditions in their inter-stellar medium (ISM)~\citep{Wang2013,Willott2015,Willott2015a}. The most luminous galaxies and quasars found at $z>6$ are expected to be highly biased tracers of the under lying dark matter distribution, forming in the most overdense regions of space. Hierarchical evolution causes these overdensities to grow with time, making it likely that these systems are progenitors of the most massive galaxies and structures we observe at any redshift \citep{Chiang2013}. This makes observations of galaxies during the EoR a crucial probe of the early evolution of these massive systems. Observations of these extreme systems and there surroundings during the EoR are key to constraining galaxy formation models.

Given the importance of understanding and characterizing overdensities in the EoR, many studies have searched for overdensities the fields surrounding quasars as possible beacons of massive halos, using various observational techniques. Early attempts leveraged the Lyman break technique to detect Lyman break galaxies (LBGs) by searching for dropouts in photometric data. This technique has yielded mixed results with some studies reporting an excess of galaxies in the fields of quasars \citep{Stiavelli2005,Zheng2006,kim2009,Husband2013}, others reporting no overedensity of galaxies \citep{Willott2005,Utsumi2010} and even \citet{kim2009} report an underdensity of LBGs in two of the five quasars fields searched. A complimentary technique is to search for Lyman alpha emitters (LAEs), which provides the advantage of searching a narrow redshift range ($\Delta z \sim 0.1$) compared the the Lyman break technique ($\Delta z \sim 1$) that may select galaxies that are physically unassociated with the quasar. This technique has produced similarly mixed results with several studies reporting no detections of LAEs in the fields of quasars at $z>6$~ \citep{Banados2013,Mazzucchelli2016, Goto2017} but \citet{Farina2017} report the detection of an LAE in close vicinity to a z $\sim$ 6.6 QSO. Combining these two techniques, \cite{Ota2018} investigated the environment of a quasar at $z=6.61$, searching for both LAEs and LBGs. They find an over-density of LBGs but an under-density of LAEs. It is likely that these techniques are probing different populations of galaxies and the authors suggest that LAEs likely reside in lower mass halos ($M_{\rm halo} \lesssim 10^{10} M_\odot$) than LBGs and thus are more easily quenched by the UV radiation field of the quasar. \citet{Champagne2018} searched for millimeter continuum sources in the fields surrounding 35 quasars at $z>6$ and found no evidence for an elevated number of sources in these fields. However, searching for galaxies using the sub/millimeter technique is only selects dusty and highly star forming galaxies (SFR $\gtrsim 100 \msunperyr$). Additionally the negative K- correction of galaxies in this regime leads to a large line of sight volume probed, thereby potentially washing out any intrinsic overdensity.

Given these varied results, there are a few explanations for why quasars may not inhabit overdense regions in the early universe. \citet{Willott2005} suggest that, due to scatter in the $\rm M_{BH}\ vs.\ M_{Halo}$ relation, quasars in the early universe may not populate as high mass halos as previously expected. This implies that quasars would not actually signpost overdense regions in the early universe, therefore finding companions would be less likely. Alternatively \citet{Utsumi2010} and \citet{Mazzucchelli2016} suggest that the lack of companions may be due to strong UV radiation from the quasar inhibiting galaxy formation, even if they reside in massive halos. However, recent studies from \citet{Trakhtenbrot2017} and \citet{McGreer2014} have found galaxies within a projected 50 kpc distance from quasars. Generally  quasars have not proved a reliable avenue to sign-post overdensities at high redshifts, and previous studies seem to indicate a complex bias (e.g. \citet*{Trainor2012}).

Attempts have also been made to search for galaxy overdensities around the most distant sub-millimeter galaxies (SMGs).Most notably HFLS3, which has a redshift of 6.34 \citep{Riechers2013}, and is one of the most extreme objects known to exist during the EoR. HFLS3 appears to be a massive starburst with a SFR of $\sim$ 2900 $\msunperyr$, with gas and dust masses of $ ~1 \times 10^{11}$ $\msun$ and $~ 1.3 \times 10^{9}$ $\msun$, respectively. Following its discovery, two studies were conducted to search for an excess of galaxies in the surrounding field \citep{Laporte2015, Robson2014}. \citet{Robson2014} searched the field around HFLS3 with SCUBA2  at 450 $\mu m$ and 850 $\mu m$ wavelengths. They found no evidence for an excess  of luminous sub-mm emitters (with implied L$_{\rm IR}>5\times10^{12}$) on a scale of 1.5 Mpc around HFLS3.  \citet{Laporte2015} used the \textit{Gran Telescopio Canarias} (GTC) and the \textit{Hubble Space Telescope} (HST) to search for an excess of LBGs in the same field. Even at the lower SFRs probed by the Lyman break technique, they do not find any significant evidence that HFLS3 is a member of a proto-cluster. 

While these results appear somewhat at odds with expectations, the studies described above suffer due to sensitivity limits and shortcomings of the selection techniques used. Optical selection of LBGs is difficult during the EoR due to the faintness of galaxies at $z>6$. Similarly, single dish sub-mm observations only select galaxies with high SFRs ($\gtrsim 100\ \rm M_\odot / yr$) and will likely miss lower mass galaxies detected through other methods. Even using more sensitive large interferometers, like the \textit{Atacama Large Millimeter Array} (ALMA), it has been shown that many UV selected galaxies are undetected in the FIR continuum~\citep{Bouwens2016}. Using ALMA to instead observe ionized carbon emission lines is a complimentary method to detecting galaxies during the EoR. Carbon has one of the lowest ionization energies of the elements that are abundant in the early universe. Due to the fine structure of ionized carbon, [CII], it is excited at 91K and then decays through the $^2 P_{3/2} \rightarrow ^2 P_{1/2}$ transition, which emits a photon at 157.7 $\mu m$. [CII] is one of the brightest emission lines in star-forming galaxies and is a major cooling mechanism in the ISM. Recent works have shown that it is possible to study [CII]  emission in high-redshift galaxies using ALMA. \citet{Capak2015} and \citet{Willott2015} studied the FIR and dust properties of galaxies using ALMA, while many studies have probed the [CII] and dust  of the host galaxies of $z\sim6$ quasar \citep{Wang2013,Willott2013,Decarli2018}.

There are some downsides to using the [CII] emission line to search for companion galaxies. The procedure used to identify candidate sources by performing a blind search of three dimensional data cubes leads to many independent measurements which could produce a high rate of false positives \citep{Aravena2016}. Additionally, different emission lines originating from galaxies at lower redshifts can be confused with [CII] at $z\sim6$. Specifically , the CO rotational lines corresponding to J $ = 3-5$ originating at $z\sim0.5-2$ appear at the same observed frequency as [CII] at $z\sim6$. \cite{Decarli2017} search for [CII] emitting companions around 25 quasars at $z>6$. They find four companions at high significance ($>7\sigma$) with L$_{\rm [CII]} > 10^9\, \lsun$. This is orders of magnitude more sources than expected given the volume probed by the ALMA observations. Therefore, the authors conclude that bright, high redshift quasars provide beacons of dark matter overdensities in the early universe.

In this paper we investigate the hypothesis that the environments of extreme objects at $z>6$ should possess overdensities of galaxies by performing a sensitive search for companions around quasars using [CII] emission lines. In Section~\ref{sec:ALMA} we define our sample ALMA fields and develop a method to search for robust [CII] line-emitting companions around previously observed extreme objects at $z\sim6$. In Section~\ref{sec:sim} we describe the results of a similar analysis performed on a simulated galaxy sample from the \citet{Hayward2013} (H13) mock galaxy catalog. Finally, the results as a whole are discussed and summarized in Sections~\ref{sec:discussion} and ~\ref{sec:conclusion} respectively. Throughout this study we assume a $\Lambda$CDM cosmology with parameters h$=0.7$~Mpc$^{-1}$, $\Omega_{\Lambda} = 0.73$ and $\Omega_{M} = 0.27$~\citep{Planck2014}.

\section{ALMA Observations}
\label{sec:ALMA}
\subsection{Sample and Methods}
\label{sec:meth}
Our sample consists of deep $\sim$\,1.2 mm ALMA observations (Band-6) of five luminous objects at $z>6$.  
We use observations of two starbursts, CLM1 and WMH5 \citep{Willott2015a}, and two quasars, CFHQSJ0210-0546 and J2329-0301  \citep{Willott2013}, as well as the data retrieved from the archive for an additional quasar J054-0005~\citet{Wang2013}. There is archival data for other quasars observed in the \citet{Wang2013} study, however the RMS noise is larger ($\rm RMS > 0.5 mJy/channel$) and a larger spectral resolution of $\sim 80 \rm MHz$, thus it is not possible to detect companions within the luminosity range of interest to this study. For this reason we have chosen not to include those fields in this study. For the first four data cubes \citep{Willott2013,Willott2015a}, we analyze the full $\sim$ 8\,GHz from the four base bands, two centred on the extreme object, and two spaced $\sim15$\,GHz away (in the upper sideband). In the archival data cubes from \citet{Wang2013}, we were only able to retrieve the 2\,GHz baseband containing the quasar itself and thus have less continuum sensitivity and frequency bandwidth to search for companions. All of this data was obtained between 2012 and 2014, and we refer the reader to the papers cited for full information about observing strategies. The raw data from the archive was re-imaged using the Common Astronomy Software Applications package (CASA v. 4.2.2, \citet{Mcmullin2007}) task \texttt{clean} using the parameters suggested in the \texttt{ScriptforImaging.py} provided by the joint ALMA observatory along with the raw data. Once imaged, further analysis of the data cubes was performed with python, relying on the \texttt{SpectralCube}\footnote{https://spectral-cube.readthedocs.io/en/latest/} python package. 

The typical beam size of the observations is $\sim 0.6^{\prime \prime}$. Given the sizes of known [CII] emitters at $z\sim6$ are known to be $\lesssim 1^{\prime \prime}$~\citep{Capak2015,Decarli2017}, we do not expect any companions to be significantly spatially resolved. Our approach to selecting sources based on the peak flux recovers all [CII] emitters found in a survey of galaxies at $z$=4.3 \citep{Miller2018}. The sensitivity of these archival observations vary by a factor of $2$ (listed in Table~\ref{table:fields}) but are on average deep enough to detect sources down to a $5\sigma$ detection threshold of L$_{\rm [CII]} \approx 10^8 \lsun$ at $z=6$ for a Gaussian line profile with FHWM$ = 150\, \rm km\, s^{-1}$.

\begin{table*}
	\centering
	\caption{Properties of fields searched for companions}
	\begin{tabular}{l | l l l l}
		Source Targeted & Frequency Coverage (GHz) & RMS Noise\textsuperscript{*} (mJy) & Beam size & Reference \\ \hline
		CLM-1 & 249.3 - 252.9 , 264.3 - 267.9 & 0.18 & $0.5^{\prime \prime} \times 0.5^{\prime \prime}$ &  \citet{Willott2015a} \\
		WMH-5 & 253.0 - 256.8 , 268.0 - 271.8 & 0.22 & $0.5^{\prime \prime} \times 0.5^{\prime \prime}$ & \citet{Willott2015a} \\
		J0210-0456 & 254.7 - 256.2 & 0.34 & $0.79^{\prime \prime} \times 0.5^{\prime \prime}$ &  \citet{Willott2013} \\
		J2329-0301 & 255.4 - 257.1 & 0.25 & $0.73^{\prime \prime} \times 0.61^{\prime \prime}$ &  \citet{Willott2013} \\
		J2054-0005 & 269.2 - 270.9 & 0.38 & $0.57^{\prime \prime} \times 0.51^{\prime \prime}$ &  \citet{Wang2013} \\ 
		\hline
		\multicolumn{4}{l}{\textsuperscript{*}\footnotesize{Per 15 MHz channel}}\\
	\end{tabular}
	\label{table:fields}
\end{table*}

To search for line candidates in the ALMA data cubes we developed a blind search algorithm. First, the entire cube was searched to find all points in the cube which exhibited a flux greater than 3$\times$ the rms noise in a single 15-MHz channel (typically 0.75 mJy beam$^{-1}$). With these positions recorded, the same positions in neighboring frequency slices were searched. If four surrounding channels (a minimal physical line width of $\sim50$ km~s$^{-1}$) had fluxes greater than 2$\times$ the rms noise (typically 0.5~mJy beam$^{-1}$), the source was deemed a possible line candidate. The significance of these candidates was then investigated. The velocity FWHM of the candidate was measured by fitting a Gaussian and a moment 0 map was constructed using the channels contained within the FWHM of the candidate. The signal to noise ratio (S/N) is calculated by dividing the integrated flux of the line by the average RMS of the moment 0 map encompassing the FWHM. In this process we ensure to mask the region of the map containing the primary target. Any object with a S/N ratio greater than 5 that lies within the FWHM of the primary beam is deemed a possible candidate. The possible candidates are inspected by hand to ensure they show Gaussian-like line profiles. Four objects were identified by the algorithm with S/N ratios greater than 5 and all passed the inspection. These objects are presented as the companions discussed in the results section below. To further test the algorithm, we lowered the S/N cutoff to 4, identifying a further 16 candidates. These objects generally exhibited lower FHWM and peak fluxes than the ${\rm S/N} > 5$ sources. We ran additional tests on the full sample of ${\rm S/N} > 4$ lines to test their purity, as described in the subsection below.

\subsubsection{Purity of sample}
Although our line candidates have a S/N greater than $5$ it is still possible that they could be spurious detections due to the non - Gaussian phase noise of the interferometer or the large number of independent measurements made during our procedure ~\citep[see ][]{Hayatsu2017,Hayatsu2019}. To estimate the rate at which false positives could occur we apply our search algorithm to find negative peaks in the data. At a S/N less than 5 we find that negative peaks at the same S/N have similar distributions and properties as positive peaks but small FWHM values and are thus likely unphysical given their fluxes. However, there was only one negative peak with $S/N>5$ (in the CLM1 cube). This suggests that 1 of the 4 line candidates is a false positive. This false positive rate, of $25\% \pm 25\%$, is consistent with the statistical analysis performed by~\citet{Aravena2016} on the ASPECS field, who predict a false positive rate of $\sim 35\%$ for a S/N cutoff of 5. An additional possibility is that we are observing a peak in the dirty beam structure from the brighter target source. By analyzing the synthesized beam output of the clean function in CASA in each case, we find no strong sidelobe structures (dirty beam sidelobe peaks $\lesssim 8\%$ of the central beam) at the positions of the candidates, suggesting that this is not a concern for the strong levels that we detect our candidate [CII] emitters. Further, the significant velocity offsets of our candidates from the central sources makes it even more unlikely that they are related to the central source beam structure.

\begin{figure}
	\centering
	\includegraphics[width = 0.9\columnwidth]{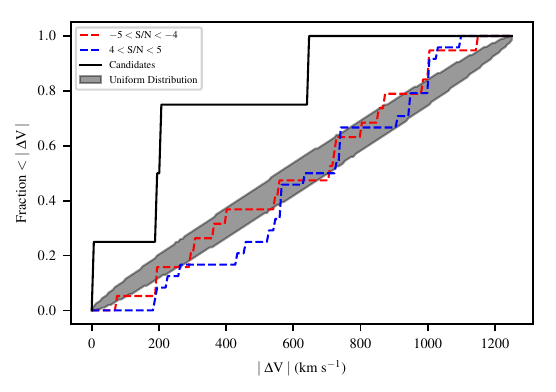}
	\caption{ This figure displays the cumulative distribution of velocity offsets between the candidate line emitters and the primary targets, along with the expected distribution if the candidates were uniformly distributed in the data cubes. The grey band shows the 5\%-95\% confidence interval of the uniform distribution, calculated through bootstrapping. Through a Monte Carlo analysis we find that the candidates are biased to a lower $\Delta$V with respect to the uniform distribution at a $2\sigma$ significance level. Also shown is the velocity offset distribution of positive and negative lines with 4 $<$ S/N $<$ 5. They appear to have a more uniform distribution, unlike the S/N $>$ 5 candidates which appear to be biased to lower velocities.}
	\label{fig:dv_dist}
\end{figure}

Even if the sources are real, we still must consider the possibility that these lines represent other transitions or species at different
redshifts, the most likely being the mid-J CO transitions. The CO (3-2), CO (4-3) and CO (5-4) transitions are observable within the same frequency band at approximate redshifts of 0.3, 0.8 and 1.2 respectively. Based on the ~\citet{Popping2016} models for the luminosity functions of the CO rotational lines at various redshifts, we naively expect to see $0.25$ sufficiently luminous low redshift CO line-emitting galaxies in the volume spanned by the 5 cubes. This calculation takes into account the rms of each field and varying rms as a function of radius due to the ALMA primary beam. As the expected number of interlopers is $<<1$ we can safely neglect this as a possibility. Even though we are probing down to low flux values where the density of interlopers is higher, the volume spanned by our cubes is small enough that the number of interlopers expected is low. The predictions for the luminosity functions from the Popping et al.\ model for the transitions and redshifts of interest agree well with current observational constraints (See results of the 1mm survey in Fig. 4 of \citet{Decarli2016}). However, these models do under-predict the number of gas rich galaxies at $z>1$~\citep{Decarli2016}. If future constraints show the predicted luminosity functions underestimate the number of bright sources for the transitions of interest, the number of expected interlopers will increase.

Figure~\ref{fig:dv_dist} displays the cumulative distribution of the velocity offsets between the candidates and the primary ALMA targets as well the expected distribution if the candidates were uniformly distributed in the data cubes. The sidebands for the CLM-1 and WMH-5 cubes are not shown in Fig.~\ref{fig:dv_dist} as the velocity offset reach $>10,000$ km s$^{-1}$, and it is difficult to show these alongside the smaller offsets. It is worth noting that there were no candidates found in these sidebands. It appears that the candidate's distribution is inconsistent with a uniform distribution and the candidates are biased towards being closer to the central galaxies. This reinforces the idea that the candidates are real galaxies that are physically associated with the primary targets as one would expect interlopers or spurious detections to be uniformly distributed in the cube. To test this we perform a Monte Carlo analysis by repeatedly sampling 4 elements from the expected distribution to calculate the probability that all 4 randomly selected elements would have a lower $\Delta$V than the maximum of the candidates (642 km/s for CLM1-A). This calculation includes the sidebands for the CLM-1 and WMH-5 cubes that are not shown in Fig.~\ref{fig:dv_dist}. After 10,000 iterations we find that 95\% of the realizations contain at least 1 of the randomly selected velocity offsets, $\Delta$V, that is larger than the maximum of the candidates. Although this is only marginally statistically significant detection ($\sim 2 \sigma$), it is consistent with  the candidates more likely appearing closer to the central galaxies with respect to a uniform  distribution in the data cubes. As the velocity offset of CLM1-A (642 km s$^{-1}$) is significantly larger than that of the next highest candidate J0210-0546-B at 205 km s$^{-1}$, we also investigate the likelihood of finding 3 candidates within 205 km s$^{-1}$. This is even less likely with $99.6$\% of the realizations containing at least 1 of 3 candidates with $\Delta V >$ 205 km s$^{-1}$, a result that is statically significant at the $\sim 3 \sigma$ level. This does not necessarily confirm the reality of our sources but simply that their $\Delta V$ distribution is inconsistent with being uniformly distributed within the cubes.

\subsubsection{Voxel Flux Distribution}
\begin{figure*}
	\centering
	\includegraphics[width = 0.9\textwidth]{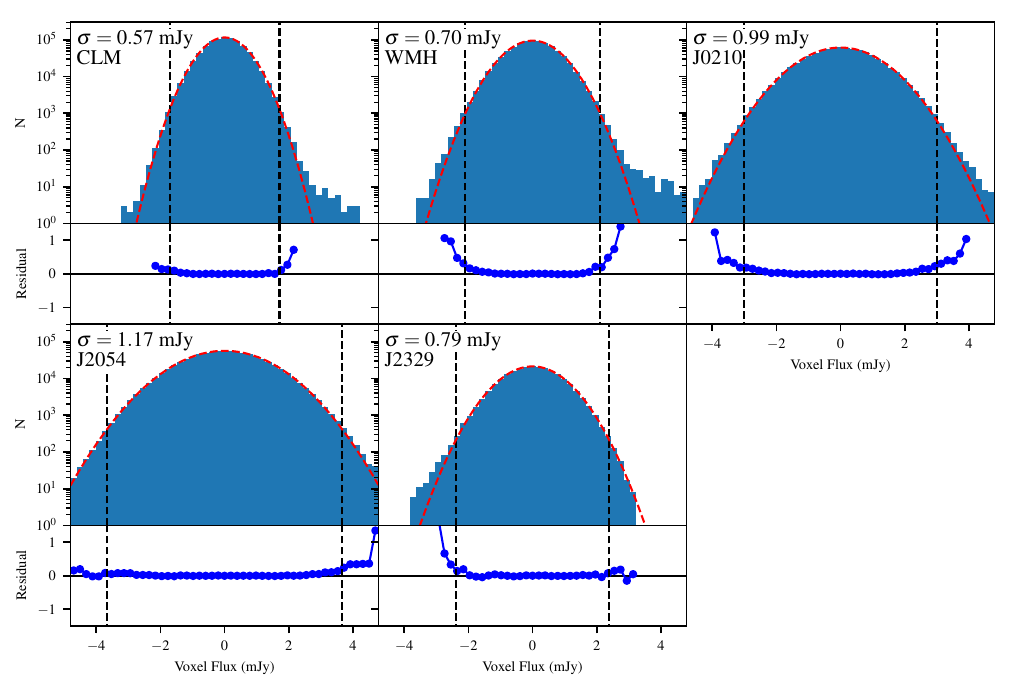}
	\caption{The top panel for each field displays the distribution of 60 MHz voxel fluxes. A Gaussian fit is shown in the red line. The ratio of the data to the Gaussian fit is displayed in the bottom panel. The vertical dotted lines shows the $\pm 3 \sigma$ range. We see the distribution of voxels in each field is well fit by a Gaussian over the $\pm 3 \sigma$ range, however at large positive and negative fluxes the distribution deviates from the Gaussian fit. The excess of voxels at large positive flux can be attributed to the sources targeted in each field but the voxels with large negative flux are likely caused by correlated noise spikes.}
	\label{fig:voxel}
\end{figure*}

In Figure~\ref{fig:voxel} we further examine the noise properties of our ALMA data cubes. We plot the distribution of 60 MHz voxels (data cube pixel) flux values for each data cube, along with a Gaussian fit and the corresponding residuals.  Since these are targeted observations, and we have not masked or removed any sources, we expect the positive side of the distribution to be skewed. Therefore we focus on the distribution of voxels with negative fluxes. The distributions in each field are well represented by a Gaussian, showing residuals of less than 1 part in 50 for the $\pm 3\sigma$ range. There appears to be an excess of voxels with negative flux in the $-4 \sigma$ to $-5 \sigma range$ compared to the overall Gaussian distribution. The excess of voxels with large negative fluxes is concerning as they are likely caused by correlated, non-Gaussian noise. As these correlated noise spikes are equally likely to produce voxels with positive flux, it is possible that they could be mistaken for line emitters and produce false positives in our sample. We have verified that the non-Gaussianity is not caused by increased noise per channel near the edge of the side-band. By plotting the distribution of voxels excluding the upper and lower 25\% of frequency slices in each cube we observe a similar excess of voxels outside the $\pm 3 \sigma$ range.

\subsection{Results}
\begin{figure*}
\centering

\subfigure[CLM1-A]{\label{fig:CLM1-A}\includegraphics[width=0.4\textwidth]{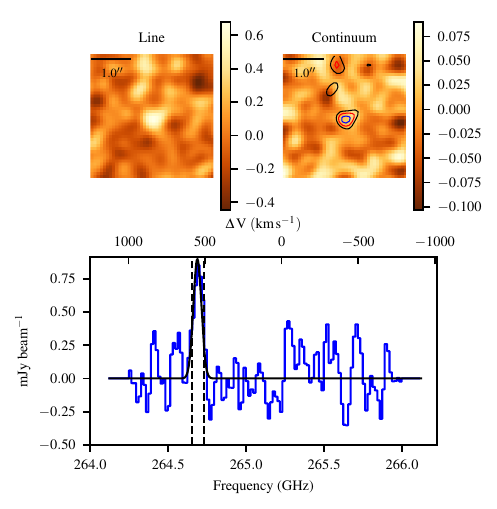}}
\subfigure[CFHQS J0210-A]{\label{fig:CHFQS-A}\includegraphics[width = 0.4\textwidth]{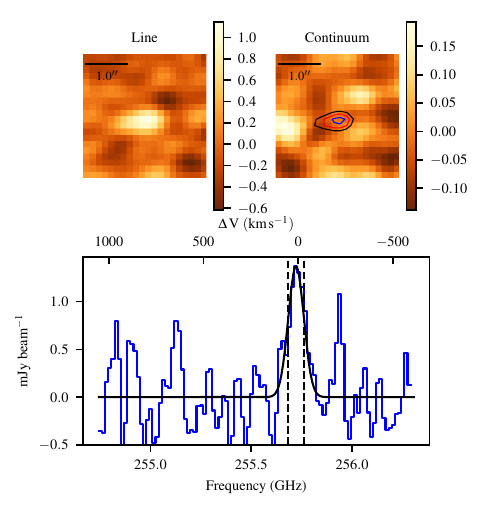}}
\\
\subfigure[CFHQS J0210-B]{\label{fig:CHFQS-B}\includegraphics[width =0.4\textwidth]{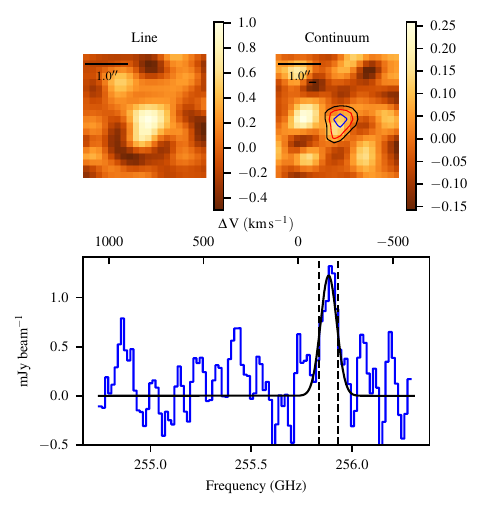}}

\caption{The four line candidates found by the blind search of the deep ALMA cubes are shown here. A Gaussian fit the 1-D spectra in black, with the vertical dotted lines denoting the FWHM of the line. The bottom axis shows the frequency of observations while the top displays the velocity offset from the primary ALMA target. We show the line flux, calculated using only the frequency slices within the FWHM of the line, as well as the continuum flux. Contours on the continuum images represent 0.9 (blue), 0.7 (red) and 0.5 (black) times the peak flux in the corresponding line channel.  a) A candidate found near the UV luminous LBG CLM1. ALMA data was originally taken by and analyzed in \citet{Willott2015a}. b) A candidate found near the Quasar J0210-0456, originally analyzed by \citet{Willott2013}. c) A second candidate found near the quasar J20210-0456. }
\label{fig:candidates}
\end{figure*}

Upon inspection we notice that one of the four candidates found by applying our search algorithm to the ALMA datasets is the source WMH5-B, previously discussed in~\citet{Willott2015a}. Willott et al. conclude that it is likely an on going merger with the more massive central galaxy WMH5-A. They conclude it cannot be classified as two distinct sources and therefore we do not consider WMH5-B for the following analysis. 

Figure~\ref{fig:candidates} displays three new line candidates used in the following analysis. The  1D spectrum, continuum and line maps for each  line candidate are shown. The channel map is extracted using the FWHM of the given line profile. The continuum map for each is constructed by using the frequency band containing the [CII] line, making sure not to include the frequency slices containing the line itself. These values are listed in Table~\ref{table:candidates}. For the CLM1-A companion we have additional data in the neighbouring side bands. We additionally investigate these side-bands for continuum emission from CLM1-A but again do not find a significant detection with the $S/N < 1$. A Gaussian function was fit to each line in order to extract a redshift  as well as the integrated flux and FWHM of the line. Observed properties of the 3 candidates are shown in Table ~\ref{table:candidates}.

\begin{table*}
\begin{center}
\caption{ Displaying properties of the 3 new line candidates found by using the blind search algorithm described in Section~\ref{sec:meth}.}
\begin{tabular}{ l| l l l }
Source Name & CLM1-A& CFHQSJ0210-0546-A & CFHQSJ0210-0546-B \\ \hline
RA (J2000) & 2:28:02.970 & 2:10:13.883 & 2:10:13.501  \\ 
DEC (J2000) & -4:16:11.74 & -4:56:22.86 & -4:56:19.26  \\
$z_{\rm [CII]}$ & 6.180 & 6.432 & 6.427 \\
$\Delta v$ (km s$^{-1}$)) & 642 & -2 & -191 \\
Proj. Sep. (kpc) & 37 & 58 & 27\\
Peak Flux (mJy) & 0.861 & 1.19 &  1.320  \\
Integrated Line  Flux (Jy km s$^{-1}$)  & 0.07 $\pm$ 0.01 & 0.16 $\pm$ 0.03 &  0.15 $\pm$ 0.03   \\ 
FWHM (km s$^{-1}$) & 75 $\pm$ 14  & 113  $\pm$ 22 & 118  $\pm$ 24   \\ 
L$_{\rm [CII]}$ ($10^8$ L$_{\odot}$) & 0.7 $\pm$ 0.2  & 1.8  $\pm$ 0.5 & 1.6  $\pm$ 0.4  \\ 
Line SNR & 5.01 & 5.12 & 5.04 \\ 
Continuum SNR & 0.77 & 0.42 & 2.27 \\ \hline
\end{tabular}
\end{center}
\label{table:candidates}
\end{table*}

Physical properties of the line candidates are listed in Table~\ref{table:candidates}. The FWHM values of the 4 candidate lines range from 75 to 189 km s$^{-1}$, the line luminosity, L$_{\rm [CII]}$,  ranges from $7 \times 10^7 \lsun$ to $2.5 \times 10^8\lsun$ (corresponding to a range of integrated line fluxes of 0.07 -- 0.25 Jy km s$^{-1}$) and only one of our line candidates are detected in the continuum at a S/N $> 2$, with the remaining two showing $S/N< 1$ continuum maps.

\begin{figure}
	\centering
	\includegraphics[width = 0.9\columnwidth]{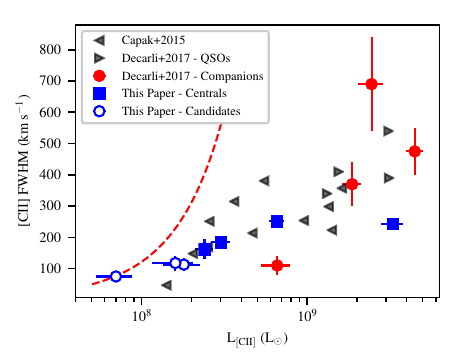}
	\caption{This figure displays our line candidates compared to the \citet{Capak2015} sample of $z\sim5$ LBGs and \citet{Decarli2017} sample of quasars and associated companions in the [CII] FWHM vs L$_{\rm [CII]}$ plane. The red line shows, for a given FWHM value the minimum luminosity needed to achieve a SNR of 5. This is based on a Gaussian line profile and typical noise in our cubes (RMS$\sim$0.25 mJy). Our line candidates and primary galaxies follow a similar distribution to the previously observed galaxies, with our candidates extending an apparent relation to slightly lower L$_{\rm [CII]}$ and FWHM, but lying significantly above our minimum detection threshold line.}
	\label{fig:l_fwhm}
\end{figure}

The observed properties of the candidate line emitters and primary galaxies are compared to previous detections of [CII] emission in high-redshift galaxies in Figure~\ref{fig:l_fwhm}. [CII] FWHM vs L$_{\rm [CII]}$ is plotted for the candidates and central sources in this study, detections of [CII] in $z\sim5$ LBGs from \citet{Capak2015} and the companions and central QSOs discussed in \citet{Decarli2017}. The dotted line shows, based on our search algorithm, the minimum luminosity needed to reach a S/N of 5 for a given FWHM value. This assumes a Gaussian line profile and the typical noise of our cubes (RMS $\sim$ 0.25 mJy per channel). Our candidates follow a similar distribution to the Capak et al.\ galaxies and the Decarli et al.\ companions, extending the apparent relation to slightly lower values of L$_{\rm [CII]}$ and FWHM. None of our candidates appear as outliers in the distribution of known [CII] emitters. One might expect false positive detections to have higher $L_{\rm [CII]}$ at a given FWHM and thereby trace the detection threshold more closely. Moreover the small volumes probed by our survey suggest that any companions we did find would be low luminosity, relatively close to our detection threshold.

\begin{figure}
\centering
\includegraphics[width = 0.9\columnwidth]{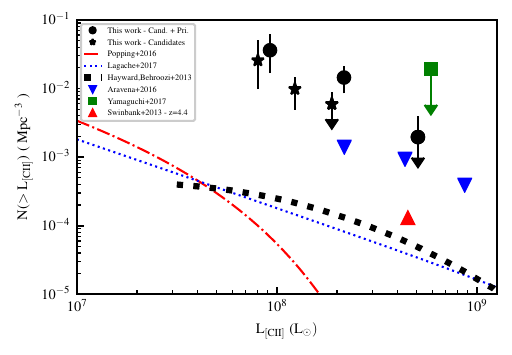}
\caption{ This plot displays various measurements and predictions for the luminosity function of [CII] emitters at $z=6$. The black squares display the density of of all sources the fields used in this study, while the black stars show the luminosity function only considering the 3 new line candidates discussed in this study. The blue triangles show recent observational constraints made be \citet{Aravena2016} at $6<z<8$ and the red triangle shows a measurement by \citet{Swinbank2012} at $z=4.4$. Observational constraints from \cite{Yamaguchi2017} are also shown in the green squares. Predictions for the $z=6$ [CII] luminosity function based on semi-analytic models discussed in \citet{Popping2016} and \cite{Lagache2017} are shown in the dotted and dot-dash lines, respectively. \citet{Hayward2013} displays a prediction from an abundance matching model combined with the empirical SFR-L$_{\rm [CII]}$ relation from \citet{DeLooze2014}. Section \ref{sec:sim} contains a full description of the Hayward et al. model. While the [CII] Luminosity function is not well constrained at L$_{\rm [CII]} \sim 10^8\ \rm L_\odot$, the luminosity function of candidates in the fields used lie at least an order of magnitude above any measurement or prediction. This suggests that luminous galaxies $z>6$ represent biased regions and therefore signpost overdensities in the early universe.}
\label{fig:alma_LF}
\end{figure}

We can calculate the luminosity function of [CII] emitters in the observed fields used in this study. We integrate the volume possible to detect a line emitting galaxy at a given luminosity by taking into account the differing noise properties of each field as well as the effect of the ALMA primary beam (assumed to be a gaussian with FWHM = $23^{\prime \prime}$ at this frequency). The redshift range covered is found through the spectral coverage of the ALMA data cubes, found in Table~\ref{table:fields}, and the known rest frame emission of the [CII] line at $157.7~\mu m$. The volume used to calculate the luminosity function is thus different for each luminosity bin. Lower luminosity galaxies cannot be detected to as large a radial distance as higher luminosity galaxies due to the effect of the primary beam on the noise amplitude in data cube, thus the volume probed is larger for high luminosity galaxies.

The luminosity function (LF) of [CII] emitters at $z\sim6$ from our study, as well as other recent measurements and predictions are displayed in Figure~\ref{fig:alma_LF}. Our data is shown both including and neglecting the primary targets of the ALMA observations. Where we only find one companion we treat our data as an upper limit. Also shown are various measurements of the field [CII] luminosity function at $z\sim6$~\citep{Swinbank2012,Aravena2016,Yamaguchi2017}. The most constraining field measurement comes from the ASPECS survey~\citet{Aravena2016}, which was calculated through a blind search for [CII] lines along with searching at the positions of known optical drop-outs. It is worth noting that this measurement along with others, are formally upper limits as the reality of all their candidates still needs to be confirmed. With future work these constraints could change significantly. Also shown are theoretical predictions of the [CII] luminosity function from previous studies~\citep{Popping2016,Lagache2017} along with a prediction from this study based on the \citet{Hayward2013} SAM (See Sec.~\ref{sec:sim}). While it appears the theoretical predictions underestimate the observations, especially the~\citet{Popping2016} prediction which drops precipitously at $ L_{\rm [CII]} > 10^8\ L_\odot$, most of the observational constraints formally represent upper limits. If indeed only one of the sources from the~\citet{Aravena2016} survey is real then the measured density of sources at $ L_{\rm [CII]} > 3\times 10^8\ L_\odot$ is only a factor of $\sim 3$ discrepant with the Lagache et al.\ prediction.

At $L_{\rm [CII]} \sim 10^8\ L_\odot$ we find our measurement of the luminosity function to be larger than any other measurement or theoretical prediction. Specifically when comparing the density of all sources (candidates presented in this paper and primary targets) at $L_{\rm [CII]} > 10^8\ L_\odot $ we find a relative overdensity of $12^{+7.5}_{-6}$ when comparing to the~\citet{Aravena2016} measurement extrapolated to lower luminosity. However, this is likely a lower limit on the overdensity, and comparing to the~\citet{Lagache2017} prediction, we find a larger relative overdensity of $158^{+104}_{-79}$. Given that these are targeted observations, the interpretation of this relative overdensity is difficult. In an attempt to correct for this we also consider the luminosity function of just the candidate companions presented in this paper. For only the companions we calculate relative overdensities of $6^{+4}_{-3}$ and $86^{+60}_{-47}$ at $L_{\rm [CII]} > 10^8\ L_\odot $ when comparing to the ~\citet{Aravena2016} observation and~\citet{Lagache2017} prediction, respectively.

\section{Simulations of [CII] emitters around the most luminous galaxies in the EoR}
\label{sec:sim}

\subsection{Mock Galaxy Catalogs}

To help interpret the results shown in Sec.~\ref{sec:ALMA} we employ mock galaxy catalogs described in \cite{Hayward2013}, where we parameterize the galaxies primarily by their observed L$_{\rm [CII]}$ and 850 $\mu$m continuum fluxes. We provide a brief description of the methodology here but refer the reader to the original paper for full details. Using a halo catalog from the \textit{Bolshoi} simulation, 8 mock light cones from $0.5<z<8$ are constructed by starting at random locations and choosing a random sight line (\citealt*{Klypin:2011}; \citealt{Behroozi:2013rockstar,Behroozi:2013}). The eight mock galaxy catalogs cover a total area of 15.7~deg$^2$ extending out to $z=8$ (1.4 $^{\circ}$ by 1.4$^{\circ}$ for each field). Stellar masses and SFRs are assigned to halos based on their mass and redshift using the functions derived in \citet*{Behroozi:2013SFH} from subhalo abundance matching\footnote{We note that these prescriptions have been recently updated in \citet{Behroozi2019} }. We then assign [CII] luminosities to galaxies in the catalog based on the power law scaling between SFR and L$_{\rm [CII]}$ empirically found in~\cite{DeLooze2014} applied with 0.42 dex of scatter, as quoted in their study. There is clearly uncertainty in the relation between [CII] luminosity and SFR during EoR yet, this simple power law scaling, empirically  derived in the local universe, has been shown to match observations fairly well~\citep{Capak2015,Vallini2015}. The 850$\mu$m flux densities (S$_{850}$) are assigned following \citet{Hayward:2013number_counts}: dust masses are computed using empirical scaling relations between gas fraction, metallicity and stellar mass. S$_{850}$ is then assigned based on a fitting function using SFR and dust mass which was derived by performing dust radiative transfer calculations on hydrodynamical simulations of galaxy mergers and isolated disk galaxies.

\subsection{Simulation Results}

\begin{figure}
\centering
\includegraphics[width = 0.9\columnwidth]{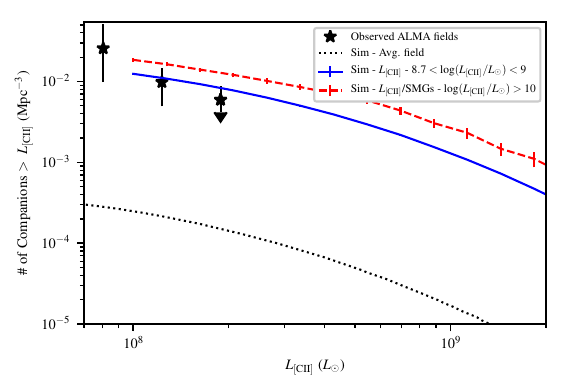}
\caption{We show the [CII] luminosity function measured from our simulation and the ALMA fields from Sec.~\ref{sec:ALMA}. The blue lines shows the luminosity function of companions surrounding simulated galaxies with $L_{\textrm{[CII]}}$ matched to the central galaxies of the ALMA observations (L$_{\textrm{[CII]}} = 0.5 - 1 \times 10^9\ \rm L_\odot$). Our observational constraints, described in Sec.~\ref{sec:ALMA} are shown as black stars. The red line shows the density of companions around the most luminous [CII] emitting galaxies in the simulation, most of which would be observable as SMGs. The black line shows the field measurement of the simulations, calculated using the entire simulation volume available. We find the simulation and observations show good agreement, reinforcing the idea that luminous galaxies at $z>6$ represent biased regions.}
\label{fig:counts_all}
\end{figure}

To directly compare the simulations with the observed counts from the ALMA data, we take the fields surrounding simulated galaxies with L$_{\textrm{[CII]}}$ matched to that of the primary ALMA targets ( L$_{\textrm{[CII]}} = 5\times10^8 - 10^9$ L$_{\odot}$ ) at $6<z<6.5$. We search the simulation around these galaxies for companions within a $15''$ radius and $dz = 0.05$, comparable to the search volume of the ALMA observations. Figure~\ref{fig:counts_all} shows the number of companions above a given [CII] luminosity for the simulated fields along with the observed ALMA fields from Section~\ref{sec:ALMA}. The simulated field counts are derived from the total $ 15.7\ \rm deg^2$ is also displayed. 

We find a relative overdensity of [CII] companions around of the matched L$_{\rm [CII]}$ sample of simulated galaxies of $50\pm 1$, consistent with the observational result. There is good agreement between the simulations and observations of the number of companions in fields surrounding galaxies of the same [CII] luminosity as the observed primary ALMA targets. The density of companions, although enhanced compared to the field measurement, follows a similar shape as the field luminosity function. It is worth noting that the simulation is incomplete at L$_{\rm [CII]} \lesssim 10^8 $ L$_\odot$ due to the minimum halo mass in the catalog.

We also show the counts surrounding the galaxies with the highest [CII] luminosities in the simulation: $\rm L_{\rm [CII]} > 10^{10} L_\odot$. These represent the most luminous simulated galaxies at this epoch and could represent the highest overdensities which are forming stars rapidly, with SFRs comparable to SMGs. This high luminosity sample of simulated galaxies consists of two populations: those with high intrinsic SFR ($>100 \msunperyr$) and those with lower intrinsic SFR that have elevated L$_{\rm [CII]}$ due to the scatter in the L$_{\rm [CII]}$-SFR relation. The former group generally has S$_{850} > 1.5$ mJy and would be detected as SMGs by current and upcoming facilities. These extremely luminous simulated galaxies contain 1.5 times the number of companions as the matched $L_{\rm [CII]}$ sample of simulated galaxies. Therefore by investigating the most luminous [CII] emitters in the simulation we find even more biased regions.

\begin{figure}
\centering
\includegraphics[width = 0.9\columnwidth]{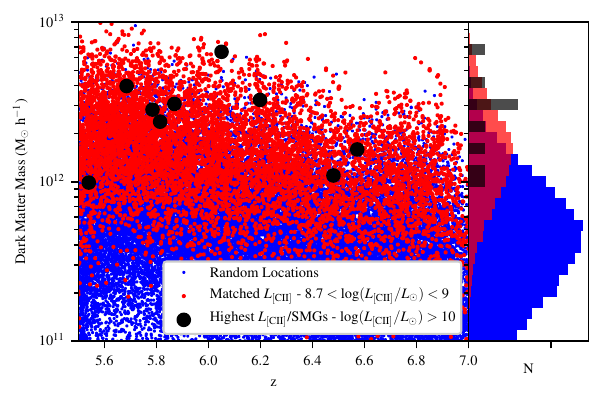}
\caption{This figure display dark matter mass in a $\sim200$ cMpc$^3$ volume centered on a given galaxy vs redshift. Dark matter mass is calculated by summing the mass of all the halos in a given region. We show  the matched [CII] luminosity sample along with the most luminous [CII] emitters or SMGs. The histogram shows the total distribution of dark matter masses for each sample between $5.5<z<7$ along with the distribution of randomly located regions. The matched L$_{\rm [CII]}$ sample often reside in overdensities, but there is a large scatter in the mass, while the highest L$_{\rm [CII]}$/SMG sample consistently reside in the most massive regions during EoR.}
\label{fig:zdm}
\end{figure}

Given that we have full information available in the mock catalog, we are able to investigate if the regions around luminous [CII] emitters during the EoR signpost peaks in the large scale matter distribution. Figure~\ref{fig:zdm} shows the total dark matter mass in a volume surrounding the matched and highest L$_{\rm [CII]}$ samples along with random locations over a redshift range of $5.5<z<7$.  The volume used to calculated the total dark matter mass is defined by $2'\times 2'$ with $dz = 0.2$ corresponding to a volume of roughly $2000\rm \ cMpc^3$. While this is larger than the volume probed by typical ALMA observations, our goal is to test the connection of [CII] emitting galaxies at $z\sim6$ to the large scale distribution of matter at this epoch. This approach was used by ~\citep{Miller2015} to investigate the bias and clustering of SMGs at $z\sim2$.

The regions surrounding  the matched L$_{\rm [CII]}$ sample of galaxies tend to have larger dark matter masses with a mean mass of $10^{12.1\pm 0.3} \rm \ M_\odot$ compared to $10^{11.5} \rm \ M_\odot$ for the random sample. However, due to the scatter some simulated galaxies in the matched luminosity range lie in relatively under-dense regions of space, while some live in the most massive regions (M $> 5 \times 10^{12} \rm \ M_\odot$). By contrast, the most extreme L$_{\rm [CII]}$ emitters consistently lie in massive overdensities with a larger mean mass of $10^{12.4 \pm 0.2} \rm \ M_\odot$ with a smaller scatter and no regions have a total mass lower than $10^{12} \rm \ M_\odot$.

\section{Discussion}
\label{sec:discussion}

We have searched for companions in archival ALMA data targeting [CII] from known $z>6$ quasars and ULIRGs. Our analysis has revealed three new and one previously known companion galaxies in the 5 fields searched. WMH5-B was previously discovered and discussed in \cite{Willott2015a}, providing validation of our method. The physical properties of our line candidates are comparable to those found by other ALMA studies targeting the [CII] line from known high-redshift galaxies \citep{Capak2015}, however the statistical analysis of the purity of our sample suggests that at least one of our candidates may be a false positive. As the faintest galaxy in our sample (CLM1-A) lies near our selection limit (Fig.~\ref{fig:l_fwhm}), with a somewhat low luminosity for its FWHM, it may not be a real galaxy. It is also possible that this offset is simply due to scatter in the L$_{\rm [CII]}$-FWHM relation. 

We are able to robustly detect lower-luminosity galaxies then other studies because the noise in these pointed deep fields is much lower than in larger area surveys. \citet{Aravena2016} perform a blind survey for $z>6$ [CII] emitters in a blank field over a similar area (7 pointings covering $\sim$ 1 arcmin$^2$) but a much larger volume due to their 7 frequency tunings over the ALMA band-6. However their average RMS of 0.56 mJy per 31.25 MHz channel is about twice that of the data we employ. They find only $\sim1-2$ line candidates which lie within the physical region of FWHM-L$_{\rm [CII]}$ region occupied by the galaxies in this study and that of \citet{Capak2015}. They note specifically that for this reason, in addition to their purity analysis, most of their candidates are probably not real. Despite the small angular size of our deep ALMA pointings, the biased regions have allowed us to uncover lower luminosity galaxies than previously found due to their increased numbers in these over-dense fields.

The lack of significant IR-continuum detections for three of our candidates is not unexpected. Known galaxies at this epoch with L$_{\rm [CII]} \sim 10^8\ \rm L_\odot$ have total IR luminosities roughly $5\times 10^{10}\ \rm L_\odot$ \citep{Capak2015}. This corresponds to an observed 1.1 mm continuum flux of roughly 30 $\mu$Jy \citep{Casey2014} while the data cubes used have a typical 3$\sigma$ detection limit of 65 $\mu$Jy. Thus our candidates would be undetected in the IR continuum if they followed these known relations. An IR luminosity exceeding $10^{11}\ \rm L_\odot$ is needed for a galaxy to be detected in the continuum with $> 3 \sigma$ confidence in the cubes used in this study. Similarly, only two of the four companions found in \cite{Decarli2017} have FIR continuum detections yet they possess [CII] luminosities over an order of magnitude larger than the companions in this study. The lack of FIR continuum detections of the companions is therefore consistent with the known relation between [CII] and FIR luminosity at $z>6$.

We may also be able to use the lack of continuum detections to rule out the possibility that our candidates are interlopers at lower redshift. To assess this possibility we consider three separate cases of CO transitions at lower redshift: CO(5-4) at z = 1.2, CO(4-3) at z = 0.8 and CO(3-2) at z = 0.3. For each case we take the average line flux of our candidates ($\sim 0.1 \rm Jy\ km/s$) and calculate the inferred gas mass from each particular line assuming a Milky Way like CO spectral line energy distribution and a conversion factor, $\alpha_{\rm_CO} = 4$. Next we infer an SFR from the gas mass using the disk model derived in \citet{Hayward2013b}, assuming a size of 1 kpc. Finally, we calculate the inferred 1.1 mm flux density, $S_{1.1 mm}$, by assuming a conversion factor of $300 \frac{\rm M_\odot / yr}{\rm mJy}$\citep{Barger2014, Hayward2013b}. We find the inferred $S_{1.1 mm}$ to be equal to $\sim 20\ \rm \mu Jy$, $\sim 10\ \rm \mu Jy$ and $\sim 5\ \rm \mu Jy$ for each case respectively. These are all under the detection threshold of the continuum data (roughly 65 $\mu$Jy), thus the lack of detection cannot rule out the possibility of interlopers. 

To further test the possibility of interlopers we perform a search of publicly available NIR and optical data. We find no sources in Spitzer IRAC (CH1 \& CH2) images at the locations of any of the three companions, however the integration times were quite short, only roughly 100 sec. We also do not find optical counterparts in either the g,r or z band of the DECaLS DR8 images \citep{Dey2019}. Using the gas masses for the three cases above and the relationship between gas mass, stellar mass and redshift shown in \citet{Hopkins2010a}, we estimate the stellar mass of the three cases of $\sim 8\times10^8\ M_\odot$, $\sim 2\times10^8\ M_\odot$ and $\sim 4\times10^7\ M_\odot$ respectively for the three cases listed above. These are very likely to be below the detection limit and therefore would go undetected in the available NIR and optical data. This is also the case if the candidates were at $z=6$. Again, the lack of NIR and optical counterparts does not constrain the possibility of our candidates being interlopers.

Since our ALMA sample is  biased to fields around extreme objects at $z>6$, we are not able to directly constrain the field luminosity function; however, we can  make predictions about the clustering and bias of galaxies at this epoch (Fig.~\ref{fig:alma_LF}). Even if the existing blank field surveys were extended to deeper flux limits comparable to our fields, we predict based on our counts/overdensity analysis that these blank field surveys would not be large enough to uncover significant numbers of fainter sources. Based on the field LF function derived in Sec~\ref{sec:sim}, a survey the size of the ASPECS survey~\citep[$\sim 1$ arcmin$^2$, see]{Aravena2016} would likely only detect one source at $L= 10^8 L_\odot$. By comparison to our simulations, we find the number counts in regions surrounding simulated galaxies with similar L$_{\rm [CII]}$ to the primary targets from the ALMA analysis show good agreement to the observed data. The factor of $ 86^{+60}_{-47} $ over density we measure from the ALMA dataset is similar to what we find in fields of extreme galaxies in the simulation, and appears to reflect the underlying matter over-density. We also note that the overall luminosity function from the simulations, shown in Fig.~\ref{fig:alma_LF}, shows fairly good agreement with another recent prediction of the [CII] field luminosity function at $z=6$ by~\citet{Lagache2017}.

These results agree with a recent study by \cite{Decarli2017}. They find roughly $16 \pm 8 \%$(4 companions for 25 targets, assuming poisson statistics) of quasars at $z>6$ host a nearby companion [CII] emitter. We find companions at a rate of $ 60 \pm 34 \% $ (3 companions for 5 targets). Our companion rate appears slightly larger, however we have adopted a lower significance and luminosity threshold (S/N$= 5$ and $\sim10^8\, \rm L_\odot$ respectively). Decarli et al. adopt a stringent 7$\sigma$ cut and therefore only find companions with L$_{\rm [CII]} > 10^9\, \rm L_\odot$. Given that we find no sources in this luminosity range, the $1 \sigma$ upper limit on our companion rate is $36\%$ (or 1.8 out of 5 fields) at L$_{\rm [CII]} > 10^9\, \rm L_\odot$, consistent with the findings of Decarli et al.

Additionally, we have shown, through the use of the simulations, that luminous [CII] emitters in the EoR not only possess an excess of companions compared to random fields but also represent overdensities in the large scale matter distribution. The simulations inevitably have some limitations, and the apparent agreement with our ALMA observations should be measured with these caveats. In the mock galaxy catalogs, only star forming galaxies parameterized by their far-IR/sub-mm emission are adopted in this realization and [CII] luminosity is assigned solely based on SFR. The quasar-phase of galaxies and the growth of the super massive black holes is not specifically treated in this implementation. Thus the connection to our three quasar fields is not entirely well motivated, although the star-forming and quasar phases have often been shown to be tightly linked (e.g., \citet{Harrison2012b,Harrison2012c}). 

\section{Conclusion}
\label{sec:conclusion}
We  present a search for companion [CII] emitters around known luminous sources during the EoR.
Using ALMA to observe [CII] emission with ALMA allows us to overcome shortcomings of other similar studies
trying to observe overdensities at $z$ $>$6 around rare and extreme sources. We develop an algorithm to search for companion [CII] line emitters in deep band-6 ALMA data of previously observed luminous galaxies and quasars. A similar analysis is then performed on a mock-galaxy catalog to put the ALMA results in context. The major results are as follows:  

\begin{itemize}
	\item We find 3 new candidate companions from our blind search of deep ALMA data of known
	luminous galaxies and quasars. All candidates display a [CII] line SNR of greater than 5,
	and lie within a projected radius of 60 kpc and 650 km s$^{-1}$ supporting the idea that they are 
	physically associated to the central galaxies.
	\item The 3 candidates display similar physical properties to previously
	studied galaxies during the EoR. We find the same L$_{\rm [CII]}$ vs. [CII] 
	FWHM relation observed in \cite{Capak2015} and \cite{Decarli2017} extended to lower luminosity values. 
	\item By calculating the luminosity function of the central galaxies and 
	the candidates we quantify the over density. These luminous galaxies represent highly biased
	regions during the EoR. Even though there are few constraints on the luminosity function
	of [CII] emitters at $z>6$ our fields show an relative overdensity of companions of at least $6^{+4}_{-3}$, when comparing to observational constraints from ~\citet{Aravena2016}, and $86^{+60}_{-47}$ when comparing to the prediction from~\citet{Lagache2017}.
	\item By performing a similar analysis on a mock galaxy catalog we find a comparable results
	to the analysis of the ALMA fields. Matching the $L_{\rm [CII]}$ of the extreme sources in the 
	simulation to the primary targets of the ALMA observations, we find a similar over-density to the field population in 
	the regions surrounding the simulated luminous [CII] emitters. Furthermore, the most luminous simulated [CII] emitting galaxies (L$_{\rm [CII]} > 10^{10} \rm L_\odot$) host even more companions, by a factor of $\sim 1.5$.
	\item By investigating the matter distribution around these sources in the simulation we find that the luminous [CII] emitters
	during EOR reside in overdense regions of space. This confirms that these simulated galaxies not only possess an excess of [CII] emitting companions in their vicinity but also signpost peaks in the large scale matter distribution.

\end{itemize}
\acknowledgments
We thank the anonymous referees, whose suggestions greatly improved the manuscript. This  paper  makes  use  of  the  following  ALMA  data: ADS/JAO.ALMA\#2013.1.00815.S. , ADS/JAO.ALMA\#2011.0.00243.S and ADS/JAO.ALMA\#2011.0.00206.S . ALMA is a partnership  of  ESO  (representing  its  member  states), NSF (USA) and NINS (Japan), together with NRC (Canada) and NSC and ASIAA (Taiwan), in cooperation with the Republic of Chile. The Joint ALMA Observatory is operated by ESO, AUI/NRAO and NAOJ. The National Radio Astronomy Observatory is a facility of the National Science  Foundation  operated  under  cooperative  agreement by Associated Universities, Inc. TBM would like to thank the Killam Trust and the Gruber Foundation for support. SCC acknowledges the Killam Trust, NSERC and CFI for support. Support for PB through program number HST-HF2-51353.001-A was provided by NASA through a Hubble Fellowship grant from the Space Telescope Science Institute, which is operated by the Association of Universities for Research in Astronomy, Incorporated, under NASA contract NAS5-26555. The Flatiron Institute is supported by the Simons Foundation.
\\ 

\bibliography{combined}

\end{document}